\documentstyle[12pt]{article}
\newtheorem{definition}{Definition}
\newtheorem{theorem}{Theorem}
\newtheorem{proposition}{Proposition}
\newtheorem{corollary}{Corollary}
\newtheorem{lemma}{Lemma}
\newcounter{c_one}

\newcommand{\be}{\begin{equation}}
\newcommand{\ee}{\end{equation}}
\newcommand{\lb}{\label}
\begin{document}
%
\begin{titlepage}
\begin{flushright}
ZU-TH 23/93
\end{flushright}
\vspace{3 ex}
\begin{center}
\vfill
{\LARGE\bf Selfgravitating Yang-Mills solitons and their Chern-Simons
numbers}
\vfill
{\bf Othmar Brodbeck and Norbert Straumann}
\vskip 0.5cm
Institute for Theoretical Physics\\Univercity of Z\"urich\\
Winterthurerstrasse 190, CH-8057 Z\"urich
\end{center}
\vfill
\begin{quote}
We present a classification of the possible regular, spherically
symmetric solutions of the Einstein-Yang-Mills system which is based on
a bundle theoretical analysis for arbitrary gauge groups. It is shown
that such solitons must be of magnetic type, at least if the magnetic
Yang-Mills charge vanishes. Explicit expressions for the Chern-Simons
numbers of these selfgravitating Yang-Mills solitons are derived, which
involve only properties of irreducible root systems and some
information about the asymptotics of the solutions. It turns out, as an
example, that the Chern-Simons numbers are always half-integers or
integers for the gauge groups $SU(n)$. Possible physical implications
of these results, which are based on analogies with the unstable
sphaleron solution of the electroweak theory, are briefly indicated.
\end{quote}
\vfill
\end{titlepage}
\section{Introduction}
Already at the time when some of us showed that the remarkable
Bartnik-McKinnon (BK) solutions \cite{bartnik} of the
Einstein-Yang-Mills (EYM) system are unstable
\cite{straumann,zhou,zhou2}, we wondered whether these solitons might
retain some physical interest, in spite of their instability. This is,
after all, the case for the sphaleron \cite{manton,klinkhamer}, which
is an unstable static solution of the classical equations of the
bosonic sector of the electroweak theory, and which is believed to play
an important role in non-perturbative baryon violating processes at
high temperatures \cite{kuzmin,arnold,ringwald,shaposhnikov}.

As a physical motivation to the present mathematical investigations we
recall that the baryon and lepton currents of the standard model have
anomalies which are proportional to a second order characteristic
density of the gauge fields. This, together with the non-trivial vacuum
structure in non-Abelian gauge theories leads to baryon and lepton
number violations at the quantum level. Classically, the infinite set
of zero energy states are classified by integer Chern-Simons numbers
$N_{CS}$, and states of different $N_{CS}$ differ also in their baryon
number. Topologically distinct vacua are separated by a potential
barrier whose minimal height is given by the energy of the spaleron.
This static solution of the classical bosonic equations has a single
unstable mode and its Chern-Simons number $1/2$ is half way between
different vacua whose topological number changes by $1$ or $-1$. It is
now widely accepted, that at high temperatures there are rapid
transitions between different vacua whose rate is proportional to the
Boltzmann factor determined by the spaleron mass. (For a short recent
review see, e.g., \cite{shaposhnikov}.)

There is a similarity between the sphaleron and the BK solution of the
EYM equations which was used by Galt'sov and Volkov \cite{galt'sov} to
give another demonstration of the instability of the BK soution. It is
conceivable that these unstable classical solutions may also be
responsible for fast baryon and lepton violating processes at extremely
high temperatures. Since the energies of the EYM solitons are set by
the Planck scale, we enter the domain of quantum gravity and therefore
do not want to become more specific. Such considerations were, however,
at the origin of our attempt to determine the Chern-Simons numbers of
regular solutions of the EYM equations for an arbitrary gauge group. In
addition, we were hoping that we might then be able to give a general
instability proof of all these solutions.

For the BK solution of $SU(2)$ the Chern-Simons number is equal to
$1/2$, as was shown recently by Moss and Wray \cite{moss} in an
explicit calculation. We shall present in this paper a thorough
analysis that is heavily based on our previous work \cite{brodbeck},
which implies in particular that the Chern-Simons numbers of EYM
solitons are always half-interger or integer for the groups $SU(n)$.
The final expressions for $N_{CS}$ involve only properties of
irreducible root systems and some information about the asymptotics of
the soliton solutions, and can easily been worked out for any
particular case.

In sect.\  2 we present a bundle theoretical classification for all
spherically symmetric EYM solitons, which implies that the possible
principal bundles are in one-to-one correspondence with a few points of
the integral lattice of the gauge group which have to be also in the
(closed) fundamental Weyl chamber. This will lead us in sect.\  3 to a
useful description of the gauge fields which we have derived in detail
already in \cite{brodbeck}. In sect.\  4 we show that regular static
and asymptotically flat solutions of the EYM equations must be of
magnetic type, at least if the magnetic YM charge of the soliton
vanishes. (This assumption is probably not necessary, but so far we
were not able to get rid of it for a general gauge group.)
The proof of this is based on an adapted form of the field equations
and makes also use of some specific group theoretical facts which are
derived in an Appendix. Before we can compute the Chern-Simons numbers,
we must construct a useful global gauge.
(A global gauge always exists since all pricipal bundles for any
compact connected gauge group over ${\cal R}^3$ or the compactified
space $S^3$ are trivial.) This construction is presented in sect.\ 5
and is then used in the next section, in which we arrive at explicit
formulae for $N_{CS}$. In a first step we derive a remarkably simple
general expression (eq.\ (\ref{ns57})), which is then further reduced
with the result collected in eqs.\ (\ref{ns70})-(\ref{ns72}). These
explicit expressions are illustrated for $SU(n)$, but can be worked out
similarly for any other gauge group. Several steps and detailed proofs
are deferred to Appendix A. In Appendix B we give further information
related to the bundle classification for spherically symmetric
solitons.
\section{Bundle classification for EYM solitons}
We begin this section by recalling results of our previous study
\cite{brodbeck} of the classification of principal bundles $P(M,G)$
over a space-time manifold $M$ for a compact connected gauge group $G$
which admit the symmetry group $K=SO(3)$ or $SU(2)$, acting by bundle
automorphisms.

We fix a maximal torus $T$ of $G$ with the corresponding integral
lattice $I$ (= kernel of the exponential map restricted to the Lie
algebra $LT$ of the torus). Furthermore, we choose a basis $S$ of the
root system $R$ of real roots, which defines the fundamental Weyl
chamber
\be
K(S)=\{\, H\in LT \mid \alpha (H)>0 \,\mbox{ \rm for all } \alpha\in S
\,\}. \lb{ns1}
\ee
Let us first consider the subbundle over a single orbit of the induced
action of $K$ on $M$. We fix a point $x_0$ on this orbit. Its isotropy
group $K_{x_0}$ maps, of course, the fiber $\pi^{-1}(x_0)$ into
itself.
Therefore, if we chose a point $u_0\in \pi^{-1}(x_0)$ and act on this
with $k\in K_{x_0}$ there exists an element $\lambda(k)\in G$ such that
$k\cdot u_0=u_0\cdot\lambda(k)$. It is easy to see that $\lambda\colon
K_{x_0}\rightarrow G$ is a group homomorphism. Changing $u_0$ to
another point on the same fiber leads to a conjugate homomorphism and
the same is true for a $K\,$-isomorphic bundle. We have thus a map
which associates to equivalence classes of principal $G\,$-bundles
admitting a fiber transitive $K\,$-action a conjugacy class of
homomorphisms $\lambda\colon\;K_{x_0}\rightarrow G$. It has been shown
in \cite{harnad} that this is a bijection.

The orbits of the induced action of $K$ by isometries on $M$ with
Lorentz metric $g$ will generically have $U(1)$ as stabilizer. {}From
Proposition 1 of our previous work \cite{brodbeck} we know that there
is a one-to-one relation between the set $H\in I\cap\overline{K(S)}$
($\overline{K(S)}$ denotes the closure of (\ref{ns1})) and the
conjugacy classes of homomorphisms from $U(1)$ to $G$ which is given as
follows: To each $H\in I\cap\overline{K(S)}$ there corresponds the
class belonging to the homomorphism $\lambda$ which is determined by
$L\lambda(2\pi i) = H$. ($L\lambda$ denotes the induced homomorphism of
the Lie algebras.)

At least in a neighbourhood of the origin of a spherically symmetric
EYM soliton $M$ is foliated by orbits which are 2-spheres (we leave out
the origin):
\be
M=\tilde M \times S^2 \qquad \mbox{(locally).}
\lb{ns2}
\ee
For each orbit $(y,S^2)$, $y\in\tilde M$, the stabilizer of
$(y,\mbox{northpole})$ is the group $U(1)$ and the corresponding
conjugacy class of homomorphisms $\lambda_y\colon U(1)\rightarrow G$ is
determined by an element from the set $I\cap\overline{K(S)}$. Clearly,
these lattice points cannot jump from one orbit to the next. This
implies that we can choose $\lambda$ independent of $y$, a result which
follows also from more general considerations \cite{harnad}; it was
pointed out in \cite{jadczyk} that this follows also from the so-called
``slice theorem'' (see \cite{bredon}).

For the origin of the regular solution the stabilizer is $K$ and we
have a corresponding conjugacy class of homomorphisms from $K$ into the
gauge group. By continuity we can choose a representative
$\tilde\lambda\colon K \rightarrow G$ such that $\tilde\lambda |_{U(1)}
= \lambda$.

Thus we arrive at an important restriction for the possible bundles of
regular solutions. The classifying homomorphism $\lambda$, defined by
$L\lambda(2\pi i)=H$, has an extension to a homomorphism
$\tilde\lambda$ from $K$ to $G$. This limits the possible lattice
points $H$ very much. As an example, consider $G=SU(n)$. Then
$\tilde\lambda$ can be regarded as an $n$-dimensional unitary
representation of $K$ and is thus a direct sum from the list $D^j$,
$j=0,\ldots,(n-1)/2$ (with only integer $j$ for $K=SO(3)$). In
particular, there is only one non-trivial possibility for $G=SU(2)$ and
$K=SU(2)$ (and none if $K=SO(3)$). This is the choice which is behind
the BK solution \cite{bartnik}. For $G=SU(3)$ there are only two
possibilities, corresponding to the three-dimensional representations
$D^1$, $D^{1/2}\oplus D^0$.

In Appendix B we give a systematic discussion for arbitrary gauge
groups which is based on extensive work by Dynkin \cite{dynkin}.
\section{Spherically symmetric gauge fields}
A spherically symmetric gauge field is given by a $K$-invariant
connection form $\omega$ on $P(M,G)$. As before, we consider first the
subbundle over a single orbit. Using previous notations, we associate
to $\omega$ the linear map $\Lambda\colon LK \rightarrow LG$ defined by
\be
\Lambda(X)=\omega_{u_0}(\tilde{X}),\rlap{$\qquad X\in LK,$}
\lb{ns3}
\ee
where $\tilde{X}$ is the induced vector field on $P$ determined by the
Lie algebra element $X$. One can show \cite{kobayashi} that the linear
map $\Lambda\colon LK\rightarrow LG\/$ satisfies
\begin{eqnarray}
\Lambda |_{LK_{x_0}} &=& L\lambda,
\lb{ns4}\\
\Lambda\circ {\rm Ad}(k) &=& {\rm
Ad}\lambda(k)\circ\Lambda,\rlap{$\qquad k\in K_{x_0}.$}
\lb{ns5}
\end{eqnarray}
According to a well-known theorem of Wang \cite{wang} there is a
one-to-one relation between the $K\,$-invariant connections on $P(M,G)$
and linear maps $\Lambda\colon LK\rightarrow LG\/$ which satisfy
(\ref{ns4}) and (\ref{ns5}). The correspondence is given by
eq.{}~{(\ref{ns3}).

For the symmetry group $K=SU(2)$ we can describe these linear maps in
more detail. Let
\be
H=2\pi i \left( \begin{array}{cr}1 & 0 \\ 0&-1 \end{array} \right),
\qquad
E_+=\left( \begin{array}{cc}0 & 1 \\ 0&0 \end{array} \right),
\qquad
E_-=\left( \begin{array}{rc}0 & 0 \\ -1&0 \end{array} \right)
\lb{ns6}
\ee
be the canonical basis for the complex extension $L{\it SU}(2)_{\cal
C}$. $H$ generates the Lie algebra of the isotropy group $U(1)$.
Condition{}~(\ref{ns4}) and our previous results imply
\be
\Lambda(H)=L\lambda(H)\in I\cap\overline{K}.
\lb{ns7}
\ee
It is easy to see that the second condition{}~(\ref{ns5}) is satisfied
if
and only if $\Lambda_+$, $\Lambda_\pm :=\Lambda(E_\pm)$, is contained
in the following direct sum of root spaces $L_\alpha$:
\be
\Lambda_+ \in  \bigoplus_{\scriptstyle \alpha\in R_+ \atop
\scriptstyle\alpha(L\lambda(H))=2 } L_\alpha, \lb{ns8}
\ee
\be
L_\alpha =\{\,X\in
LG_{\cal C} \mid [H, X]=2\pi i\alpha (H)X\;\hbox{ for all } H\in LT
\,\}.
\ee
Here, $R_+$ denotes the positive roots of the set $R\/$ of real roots
of $G$. (For the normalisation of real roots we use the same convention
as in the Ref.{}~\cite{brocker}.)

{}From this we conclude that $\Lambda_\pm\neq0$ only if the classifying
homomorphism  $\lambda\colon U(1)\rightarrow G$ of $SU(2)$ symmetric
principal bundles $P(M,G)$ satisfies
\be
H_\lambda := L\lambda(2\pi i\,)\in (I\cap\overline{K})\cap
\bigl(\bigcup_{\alpha\in R_+} L_{\alpha 2}\,\bigr), \lb{ns9}
\ee
where $L_{\alpha n} =\alpha^{-1}\{n\}$, $\alpha\in R_+$, $n\in {\cal
Z}$. We recall that the union of these planes
\be
LT_s=\bigcup_{\alpha,n} L_{\alpha n}
\ee
is the Stiefel diagram of $G\/$. It consists of the inverse image of
the set of singular elements of $T$ under the exponential map and plays
an important role in the theory of compact Lie groups (see, e.g.,
Ref.{}~\cite{brocker}).

We summarize these conclusions in the following
\begin{proposition} With the notations introduced above the equivalence
classes of principal $G\,$-bundles admitting a fiber transitive $SU(2)$
action characterized by the elements $L\lambda(2\pi i)\in
I\cap\overline{K}$ have only ``Abelian'' connections with
$\Lambda(E_\pm )=0$ (see eq.{}~(\ref{ns3})) if $L\lambda(2\pi i)$ is
not
in the part $\,\bigcup_{\alpha\in R_+}L_{\alpha 2}$ of the Stiefel
diagram.
\end{proposition}
Let the decomposition of the gauge potential relative to (\ref{ns2}) be
$A=\tilde{A}+\hat{A}$. The following Proposition was proven in
\cite{brodbeck}:
\begin{proposition} One can always find a gauge such that
\be
\hat{A}=\Lambda\circ\theta,
\lb{ns11}
\ee
where $\theta$ is a pull-back of the Maurer-Cartan form of $K$ to the
homogeneous space $K/K_{x_0}$ with any cross section, and $\tilde{A}$
is a one-form on $\tilde{M}$ which is invariant under ${\rm
Ad}\,\lambda(K_{x_0})$.
\end{proposition}

It should be noted that $\Lambda$ varies now from orbit to orbit,
$\Lambda\colon\tilde{M}\times LK\rightarrow LG$, except for $\Lambda
|_{LK_{x_0}}$ which can be chosen as $L\lambda$ for a constant
$\lambda$ as noted above.

For $K=SU(2)$, $K/K_{x_0}\cong S^2$, we choose the cross section
$\sigma$ from $S^2$ to $SU(2)$ as follows:
\be
\sigma(\vartheta,\varphi)=\exp (\varphi\tau_3)\exp(\vartheta\tau_2),
\lb{ns30a}
\ee
where $\tau_k:=\sigma_k/2i$ ($k=1,2,3$) and $\vartheta$, $\varphi$ are
the standard coordinates of $S^2$.
We have
\be
\begin{array}{rcl}
\theta&=&\sigma^{-1}d\sigma\\
      &=&\tau_2\, d\vartheta
	 +(\cos \vartheta\,\tau_3-\sin\vartheta\,\tau_1)\,d\varphi,\\
\end{array}
\lb{ns31a}
\ee
and thus (\ref{ns11}) leads with $\Lambda_k:=\Lambda(\tau_k)$ to
\be
\hat{A}={\Lambda_2
\,d\vartheta}+{(\Lambda_3\cos\vartheta-\Lambda_1\sin\vartheta)\,d\varphi}.
\lb{ns12}
\ee
\section{Magnetic structure of EYM solitons}
In this section we prove that the YM-fields of a regular static and
asymptotically flat solution of the EYM equations must be of magnetic
type for any gauge group. So far we were, unfortunately, only able to
prove this under the assumption that the magnetic charge of the soliton
vanishes. (We are currently trying to generalize the arguments in Ref.
\cite{bizon} for the gauge group $SU(2)$). The proof given below relies
on the detailed form of the coupled field equations which we have
derived in \cite{brodbeck}.

For later use we specialize the relevant results of \cite{brodbeck} to
static fields. Using intrinsically defined Schwarzschild like
coordinates we can adapt the metric to (\ref{ns2}) in the form
\be
g=\tilde g + r^2 \hat g, \qquad \tilde g= -NS^2 dt^2 + N^{-1} dr^2,
\lb{ns13}
\ee
where $\hat g$ is the standard metric on $S^2$. The metric functions
$N$ and $S$ depend only on the radial Schwarzschild coordinate $r$.

{}From now on we consider only the {\it generic case\/} for which the
classifying element $H_\lambda$ in (\ref{ns9}) is in the {\it open\/}
Weyl chamber $K(S)$. The centralizer of $H_\lambda$ (and thus the
centralizer of $\Lambda_3$) is then the infinitesimal torus $LT$ (see
sect.\ V.2 of Ref.\ \cite{brocker}). The component ${\tilde A}$ of the
gauge potential is thus Abelian and we can use a temporal gauge:
\be
A=\tilde A + \hat A, \qquad \tilde A = (NS{\cal A})\,dt,
\lb{ns14}
\ee
where ${\cal A}$ satisfies $[\Lambda_3,{\cal A}]=0$ and $\hat A$ is
given by (\ref{ns12}). Introducing the quantities
\be
\check{\cal F} = -r^2 \frac{(NS{\cal A})'}{S},
\qquad
\hat {\cal F}= [\Lambda_1,\Lambda_2]-\Lambda_3
\lb{ns15}
\ee
one can write a particular component of the YM equations as
\be
2{\check{\cal F}}'
+ [\Lambda_{+},\dot{\Lambda}_{-}]
+ [\Lambda_{-},\dot{\Lambda}_{+}]
=0,
\lb{ns16}
\ee
where
\be
\dot{\Lambda}_{\pm}:=[{\cal A},\Lambda_{\pm}].
\lb{ns17}
\ee
(See eq.\ (34) of Ref.\ \cite{brodbeck}.)

We shall need also the Einstein equation
\be
m'= 4\pi G\, r^2\varrho,
\lb{ns18}
\ee
where $m(r)$ is the usual mass fraction defined by $N=1-2m/r$, and
$\varrho$ is the energy-mass density given by
\be
\varrho = \frac{N}{2r^2}
\Bigl\{
 |\dot\Lambda_{+}  |^2
+|     \Lambda_{+}'|^2
\Bigr\}
+\frac{1}{2r^4}
\Bigl\{
 |\check{\cal F}|^2
+|\hat  {\cal F}|^2
\Bigr\}.
\lb{ns19}
\ee
(See eqs. (39)-(41) in \cite{brodbeck}.) The norm of any element
$Z=X+iY$ ($X,Y\in LG$) in the complex extension $LG_{\cal C}$ is
defined by $|Z|^2=\langle X+iY,X-iY\rangle$, where $\langle
.\,,.\rangle$ denotes an Ad$G$ -invariant scalar product in $LG$.

According to (\ref{ns8}) we know that
\be
\Lambda_+ \in  \bigoplus_{\scriptstyle \alpha\in S(\lambda )} L_\alpha,
\lb{ns20}
\ee
where
\be
S(\lambda) =\{\,\alpha\in R_+ \mid \alpha(H_\lambda)=2\,\}.
\lb{ns21}
\ee
Now, since ${\cal A}$ is $LT$-valued we can split it with respect to
the decomposition
\be
LT=\langle S(\lambda)\rangle \oplus \langle S(\lambda)\rangle^\perp,
\lb{ns22}
\ee
where $\langle S(\lambda) \rangle$ denotes the linear span of
$S(\lambda)$. We emphasize that the decomposition (\ref{ns22}) is
independent of the choice of the Ad$G$ -invariant scalar product (see
Appendix A). In obvious notation we set
\be
{\cal A} = {\cal A}_\parallel + {\cal A}_\perp
\lb{ns23}
\ee
and show first, that ${\cal A}_\perp$ can be gauged away.

To prove this we consider the perpendicular component of (\ref{ns16})
and make use of the fact
\be
[\Lambda_+,\dot{\Lambda}_-]\in \langle S(\lambda)\rangle,
\lb{ns24}
\ee
which follows from (\ref{ns17}), (\ref{ns20}) and standard commutation
relations for root spaces (using the property that for $\alpha,\beta\in
S(\lambda)$, $\alpha-\beta$ is never a root; see Appendix A). This
leads to $\check{\cal F}'_\perp=0$. {}From (\ref{ns19}) one sees that
$\check {\cal F}$ must vanish at the origin and hence ${\check{\cal
F}}_\perp$ must vanish identically. Together with (\ref{ns15}) we
obtain thus
\be
NS{\cal A}_\perp = \mbox{const}.
\lb{ns25}
\ee
A gauge transformation
\be
A\rightarrow \mbox{Ad}(g)^{-1}A + g^{-1}\,dg
\ee
with $g=\exp (-NS{\cal A}_\perp t)$ does not affect $\hat A$ in
(\ref{ns14}) (because of (\ref{ns20})), but eliminates the piece
$\tilde A_\perp$.

Next we show that ${\cal A}_\parallel$ must vanish for a regular
solution. This time we take the scalar product of (\ref{ns16}) with
$NS{\cal A}_\parallel$. With (\ref{ns17}) and the Ad$G$ -invariance of
the scalar product we obtain
\be
\langle \check {\cal F}',NS{\cal
A}_\parallel\rangle+NS|\dot{\Lambda}_+|^2=0
\ee
or with (\ref{ns15})
\be
{\langle \check {\cal F},NS{\cal A}_\parallel\rangle}'
+\frac{S}{r^2}|\check {\cal F}_ \parallel |^2+NS|\dot{\Lambda}_+|^2=0.
\lb{ns26}
\ee
Integrating this gives
\be
-\langle \check {\cal F},NS{\cal A}_\parallel\rangle{\Bigl |}_0^\infty
=\int_0^\infty
\left(\frac{S}{r^2}|\check {\cal F}_\parallel
|^2+NS|\dot{\Lambda}_+|^2\right)\,dr.
\lb{ns27}
\ee
The boundary terms on the left hand side vanish: As noted above
$\check{\cal F}(0)=0$ and from (\ref{ns18}) we see that the total
energy $\lim_{r \rightarrow \infty} m(r)$ is only finite if $\lim_{r
\rightarrow \infty}r^2\varrho=0$. According to (\ref{ns19}) this is
only possible if
\be
\lim_{r \rightarrow \infty}\,[{\cal A},\Lambda_+]=0.
\lb{ns28}
\ee
Now, we shall show in the next section that $\lim_{r \rightarrow
\infty}\Lambda$ is a homomorphism of $LSU(2)$ to $LG$ if the magnetic
YM charge of the soliton vanishes. Using this and Lemma A1 in Appendix
A it follows from (\ref{ns28}) that $\lim_{r\rightarrow \infty}{\cal
A}_\parallel=0$.

Since the integrand in (\ref{ns27}) is non-negative, we conclude that
$\check{\cal F}_\parallel=0$, implying $(NS{\cal A}_\parallel)'=0$.
Since $\lim_{r \rightarrow \infty}{\cal A}_\parallel=0$ we are led to
the desired conclusion that ${\cal A}_\parallel$ is identically equal
to zero.
\section{Construction of a useful global gauge}
For the calculation of the Chern-Simons number of a static EYM soliton
it is necessary to construct an explicit global gauge. The existence of
a global gauge for any compact connected gauge group $G$ is clear,
since all principal $G$-bundles over ${\cal R}^3$ or the compactified
space $S^3$ are trivial. This is obvious for ${\cal R}^3$ because this
is a contractible space. On the other hand, the principal $G$-bundles
over $S^3$ are classified by the second homotopy group $\Pi_2(G)$ which
is trivial for any compact connected group. (That $\Pi_2(G)$ classifies
the bundles is easy to see, since a principal bundle over $S^3$ is
given by a transition function on the intersection on two hemispherical
neighbourhoods and two such transition functions lead to isomorphic
bundles, if their restrictions to the equator are in the same homotopy
class.)

{}From sect.\ 4 we know that the gauge field is purely magnetic
\be
\begin{array}{rcl}
A&=&{\Lambda_2\,d\vartheta}
   +{(\Lambda_3\cos\vartheta-\Lambda_1\sin\vartheta)\,d\varphi}\\
 &=&\Lambda (\sigma^{-1}d\sigma),\\
\end{array}
\lb{ns29}
\ee
where $\sigma$ is the cross section from $S^2$ to $SU(2)$ introduced in
eq.\ (\ref{ns30a}).

Next we show that $\Lambda_{(0)}:=\lim_{r \rightarrow 0}\Lambda (r)$ is
a homomorphism from $LK$ to $LG$. Indeed, from (\ref{ns19}) we see that
the energy density is only well-behaved if $\hat{\cal F}$ goes to zero
for $r \rightarrow 0$. A look at (\ref{ns15}) shows then that the
$\Lambda_k:=\Lambda(\tau_k)$ satisfy for $r=0$ the commutation
relations of $LSU(2)$ ($[\Lambda_1,\Lambda_2]=\Lambda_3$, cyclic).

Similarily, the total YM charge of the soliton vanishes only if
$\lim_{r \rightarrow \infty} \hat{\cal F}=0$. Again (\ref{ns15})
implies that $\Lambda_{(\infty)}:=\lim_{r \rightarrow \infty}
\Lambda(r)$ is a homomorphism $LK \rightarrow LG$.

We have thus two homomorphisms $\Lambda_{(0)}$, $\Lambda_{(\infty)}$
which extend the classifying homomorphism $L\lambda\colon
LU(1)\rightarrow LG$ introduced earlier (see sect.\ 2). In Appendix A
we show that there exists an element $Z\in LT$ such that
\be
\Lambda_{(\infty)} = \mbox{Ad}(\exp Z ) \, \Lambda_{(0)},
\lb{ns32}
\ee
at least if $H_\lambda$ in (\ref{ns9}) is in the open Weyl chamber.
(This is what we called the generic case in sect.\ 4.)

Now we are ready to transform (\ref{ns29}) to a new gauge which is
constructed such that the new potential is well defined on all of
${\cal R}^3 \cup\infty$.
In
\be
A\rightarrow\mbox{Ad}(G^{-1})A+G^{-1}dG=:\tilde A
\lb{ns33}
\ee
we choose $G$ of the form
\be
G=g\,\Pi(\sigma^{-1}),
\lb{ns34}
\ee
where $\Pi$ is the homomorphism from $SU(2)$ to the gauge group such
that $L\Pi = \Lambda_{(0)}$, and
\be
g=\exp(\chi Z)
\lb{ns35}
\ee
with $Z$ appearing in (\ref{ns32}) which we multiply with a smooth real
function $\chi$ that satisfies
\be
\chi  =  \left \{
\begin{array}{ll}
0    & \mbox{ for }r<1-\epsilon,\\
1    & \mbox{ for }r>1+\epsilon,
\end{array} \right.
\lb{ns36}
\ee
for an $\epsilon > 0$. (For clarity we note that $\sigma^{-1}$ in
(\ref{ns34}) does not denote the inverse map but
$(\sigma^{-1})(\vartheta,\varphi):=
({\sigma(\vartheta,\varphi)})^{-1}$.)

For the determination of $\tilde A$ we compute first $G^{-1}dG$. Eq.
(\ref{ns34}) gives
\be
G^{-1}dG = \mbox{Ad}(\Pi(\sigma))\,g^{-1}dg
	 \,+\, \Pi(\sigma)\,d(\Pi(\sigma^{-1})).
\lb{ns37}
\ee
The last term is easily seen to be
\be
  \Pi(\sigma)\,d(\Pi(\sigma^{-1}))
=-\mbox{Ad}(\Pi(\sigma)) \,
  L\Pi(\sigma^{-1}d\sigma).
\lb{ns38}
\ee
Using also (\ref{ns35}), eq. (\ref{ns37}) leads to
\be
G^{-1}dG=\mbox{Ad}(\Pi(\sigma))
\Bigl\{
Z\chi'dr-L\Pi(\sigma^{-1}d\sigma)
\Bigr\}.
\lb{ns39}
\ee
The first term of $\tilde A$ is according to (\ref{ns29})
\be
\mbox{Ad}(G^{-1})\,A=\mbox{Ad}(\Pi(\sigma)g^{-1})\,\Lambda(\sigma^{-1}d\sigma)
\lb{ns40}
\ee
Together we find, using also $\mbox{Ad}(g^{-1})Z=Z$ and
$L\Pi=\Lambda_{(0)}$,
\be
\tilde A =
\mbox{Ad}(\Pi(\sigma)g^{-1})
\Bigl\{
\Bigl (
\Lambda-\mbox{Ad}(g)\Lambda_{(0)}
\Bigr )
(\sigma^{-1}d\sigma)+Z\chi'dr
\Bigr\}.
\lb{ns41}
\ee
The crucial point is now that this expression is well defined both for
$r\rightarrow 0$ and $r\rightarrow\infty$. Indeed, for $r\rightarrow 0$
we have with (\ref{ns35}), (\ref{ns36}):
$\Lambda-\mbox{Ad}(g)\Lambda_{(0)}\rightarrow 0$, and for
$r\rightarrow\infty$ this quantity vanishes because of (\ref{ns32}).
Moreover, the last term in (\ref{ns41}) vanishes for $r\notin
[1-\epsilon,1+\epsilon]$. (Note, that the right hand side of
(\ref{ns41}) is  independent of the section $\sigma$, the potential
$\tilde A$ can thus be extended across the polar axis.)

The gauge pontential $\tilde A$ is thus well-defined on the
compactified three-space ${\cal R}^3 \cup \infty$, and we can now turn
to the computation of the Chern-Simons numbers.
\section{Computation of the Chern-Simons numbers for regular solitons}
Before we come to the derivation of a concise general formula of the
Chern-Simons numbers, we recall some well-known facts.

Consider a principal bundle $P(M,G)$ with a connection form $\omega$
and a bilinear Ad$G$ -invariant form $f$ on the Lie algebra $LG$. If
$\Omega$ is the curvature of $\omega$ one easily derives with the help
of the structure equation
$\Omega=d\omega+{\textstyle\frac{1}{2}}[\omega,\omega]$ the following
idendity
\be
f(\Omega,\Omega)=
d
\bigl\{
f(\omega,d\omega + {\textstyle\frac{1}{3}}[\omega,\omega])
\bigr\}.
\lb{ns42}
\ee
As a special case of the Chern-Weyl theorem $f(\Omega,\Omega)$ projects
to a (unique) closed 4-form on $M$ whose de Rham cohomology class is
independent of the connection.

The 3-form
\be
2Q(\omega)
=f(\omega,d\omega + {\textstyle\frac{1}{3}}[\omega,\omega])
=f(\omega, \Omega - {\textstyle\frac{1}{6}}[\omega,\omega])
\lb{ns43}
\ee
is, however, not horizontal, and therefore does in general not project
to the base manifold $M$. To each local section $\sigma\colon U\subset
M\rightarrow P$ we can define a {\it local \/} Chern-Simons form in
terms of $A=\sigma^{\ast}\omega$
\be
2Q(A):=f(A,dA + {\textstyle\frac{1}{3}}[A,A]).
\lb{ns44}
\ee
Under a gauge transformation one finds the following transformation law
for the local Chern-Simons form
\be
2Q(\mbox{Ad}(g^{-1})A + g^{-1}dg)
= 2Q(A)-{\textstyle\frac{1}{6}}f(\theta,[\theta,\theta])
+ d
\bigl\{
f(\mbox{Ad}(g^{-1})A,\theta)
\bigr\},
\lb{ns45}
\ee
where $\theta$ stands for the Lie algebra valued 1-form $g^{-1}dg$,
which is the pull-back of the Maurer-Cartan form on $G$ by the
transition function $g$.

For the regular solutions considered in this paper the principal $G$
-bundles are trivial and there exist thus global sections. We are only
interested in static solitons and can then consider these transition
functions as maps from compactified 3-space ${\cal R}^3 \cup
\{\infty\}\cong S^3$ to $G$. Within the class of these global sections
the integral of $Q(A)$ over $S^3$ changes according to (\ref{ns45}) by
\be
\int_{S^3}Q
\bigl(
\mbox{Ad}(g^{-1})A + g^{-1}dg
\bigr)
= \int_{S^3}Q(A)-{\textstyle\frac{1}{12}}
  \int_{S^3}f(\theta,[\theta,\theta])
\lb{ns46}
\ee
Here, the last integral has a topological meaning and is an integer if
$f$ is suitably normalized. For example, if $G=SU(2)$ this integral is
proportional to the winding number of $g\colon S^3\rightarrow
SU(2)\cong S^3$. The Chern-Simons number
\be
N_{CS}:=\int_{S^3} Q(A)
\lb{ns47}
\ee
is thus defined up to an integer if the Ad$G$ -invariant bilinear form
$f$ is properly normalized.
\subsection{A general formula for $N_{CS}$ for an arbitrary gauge
group}
Our next task is to compute $N_{CS}$ for the globally defined gauge
potential $\tilde A$ given in (\ref{ns41}). It would, however, be
complicated to start directly from (\ref{ns41}). An easier way is to
use the local transformation law (\ref{ns45}) for the gauge
transformation (\ref{ns33}) and to compute the integral of $Q(\tilde
A)$ over all of $S^3$ by a limiting process, whereby one first excludes
conical regions around the polar axis (see Fig.\ 1). The last term in
(\ref{ns45}) gives then a boundary contribution which has to be
evaluated carefully in the limit when the conical regions shrink to the
polar axis.

\begin{figure}
\vspace{10truecm}
\caption{Integration domain in the limiting procedure.}
\end{figure}
We show first that $Q(A)=0$ for $A$ given by (\ref{ns29}):
\be
A=\Lambda_2\,d\vartheta +
(\Lambda_3\cos\vartheta-\Lambda_1\sin\vartheta)\,d\varphi.
\lb{ns48}
\ee
It is easy to see that $Q(A)$ is proportional to $\langle
\Lambda_+,\Lambda_-'\rangle-\langle \Lambda_-,\Lambda_+'\rangle$ if we
choose for $f$ an Ad$G$ -invariant scalar product $\langle.\,,.\rangle$
on $LG$. Now, we can decompose $\Lambda_+$ with respect to a basis
$\{e_\alpha\}$ of the root spaces $L_\alpha$, $\alpha\in S(\lambda)$
(see eq. (\ref{ns20})):
\be
\Lambda_+ = \sum_{\alpha\in S(\lambda)} w^\alpha e_\alpha,
\lb{ns49}
\ee
where the amplitudes $w^\alpha$ are functions of $r$ alone. Using the
orthogonality properties of the root spaces we see that $Q(A)$ is
proportional to %
\[\sum_{\alpha\in S(\lambda)}
(w^\alpha{\overline{w}^\alpha}'-\overline{w}^\alpha{w^\alpha}')\]
(for a suitable normalisation of the $e_\alpha$). This expression
vanishes, however, as a consequence of the YM equations. Indeed,
eq.\ (\ref{ns33}) of Ref.\ \cite{brodbeck} reduces for a static
solution to
\be
[\Lambda_+,\Lambda_-']+[\Lambda_-,\Lambda_+']=0,
\lb{ns50}
\ee
which gives, by inserting (\ref{ns49}) and using the fact that
$S(\lambda)$ is a simple system of roots (see Appendix A)
\be
w^\alpha {\overline{w}^\alpha}'- \overline{w}^\alpha{w^\alpha}'= 0
\qquad
\mbox{for each } \alpha\in S(\lambda).
\lb{ns51}
\ee
This equation just says that $w^\alpha$ is real, up to a constant
phase, a fact that we shall use againg later on. (We could absorb this
phase by a redefinition of the base vectors $e_\alpha$.)

Next, we compute the second term in (\ref{ns45}) for $\theta=G^{-1}dG$
given in (\ref{ns39}). Using $L\Pi =\Lambda_{(0)}$ and the Ad$G$
-invariance of the scalar product, we find
\be
\langle\theta,[\theta,\theta]\rangle
= 3\,
\langle\,
\chi'Zdr\,,\,
[\Lambda_{(0)}(\sigma^{-1}d\sigma)\,,\Lambda_{(0)}(\sigma^{-1}d\sigma)]
\,\rangle.
\lb{ns52}
\ee
Since $\Lambda_{(0)}$ is a homomorphism we have
\be
[\Lambda_{(0)}(\sigma^{-1}d\sigma)\,,\Lambda_{(0)}(\sigma^{-1}d\sigma)]
=2{\Lambda_{(0)}}_3\,\mbox{vol}_{S^2}
+2{\Lambda_{(0)}}_1\cot\vartheta\,\mbox{vol}_{S^2}.
\lb{ns53}
\ee
Only the first term contributes to (\ref{ns52}) because $Z\in LT$,
giving
\be
\langle\theta,[\theta,\theta]\rangle
=6\,\langle Z,\Lambda_3\rangle \chi'\, dr\wedge \mbox{vol}_{S^2}.
\lb{ns54}
\ee
(Recall that $\Lambda_3$ is independent of $r$.) Integration gives with
(\ref{ns36})
\be
-{\textstyle\frac{1}{12}}
\int\langle\theta,[\theta,\theta]\rangle
=-2\pi\langle Z,\Lambda_3\rangle \chi {\bigr |}_0^\infty
=-2\pi\langle Z,\Lambda_3\rangle.
\lb{ns55}
\ee

Finally, we have to calculate the surface integral arising from the
last term in (\ref{ns45}). Using (\ref{ns34}) and (\ref{ns39}), we have
\begin{eqnarray}
\lefteqn{
\langle\,
\mbox{Ad}(G^{-1})\Lambda(\sigma^{-1}d\sigma)\,,G^{-1}dG
\,\rangle
}\nonumber\\
& = &
\langle\,
\Lambda(\sigma^{-1}d\sigma)\,,
\mbox{Ad}(g)\{Z\chi'dr-\Lambda_{(0)}(\sigma^{-1}d\sigma)\}
\,\rangle
\lb{ns56}\\
& = &
\langle\,
\Lambda(\sigma^{-1}d\sigma)\,,Z
\,\rangle\,\chi'\,dr
\,-\,
\langle\,
\Lambda(\sigma^{-1}d\sigma)\,,
\mbox{Ad}(g)\Lambda_{(0)}(\sigma^{-1}d\sigma)
\,\rangle.
\nonumber
\end{eqnarray}
The last term is proportional to $d\vartheta\wedge d\varphi$ and gives
thus only contributions from the small and from the large spheres in
Fig.\ 1 in the limit $r\rightarrow 0$ and $r\rightarrow\infty$,
respectively. For $r\rightarrow 0$ this becomes (see (\ref{ns35}),
(\ref{ns36})) %
\[\langle\,
\Lambda_{(0)}(\sigma^{-1}d\sigma)\,,
\Lambda_{(0)}(\sigma^{-1}d\sigma)
\,\rangle,\] %
and for $r\rightarrow\infty$ we find with (\ref{ns32}) %
\[\langle\,
		 \Lambda_{(\infty)}(\sigma^{-1}d\sigma)\,,
\mbox{Ad}(\exp Z)\Lambda_{(0)}     (\sigma^{-1}d\sigma)
\,\rangle
= \langle\,
\Lambda_{(0)}(\sigma^{-1}d\sigma)\,,
\Lambda_{(0)}(\sigma^{-1}d\sigma)
\,\rangle.\] %
Hence, the two terms cancel.

There remains the first term in (\ref{ns56}) which is equal to %
$-\langle Z,\Lambda_3\rangle \chi'\cos\vartheta\, dr\wedge d\varphi$ %
and gives the following contribution from the two conical surfaces in
Fig.\ 1 in the limit $\vartheta\rightarrow 0,\pi$:
\begin{eqnarray*}
-\frac{1}{2}\langle Z,\Lambda_3\rangle \int \chi'\cos\vartheta\,
dr\wedge d\varphi
& \rightarrow &
-\frac{1}{2}\langle Z,\Lambda_3\rangle \cdot 2\pi\cdot 2
\cdot\chi{\bigr |}_0^\infty \\
& = &
-2\pi\langle Z,\Lambda_3\rangle.
\end{eqnarray*}
Note, that this is exactly equal to the other piece (\ref{ns55}).
Together we obtain the remarkably simple result
\be
\int Q(\tilde A)= -4\pi\langle Z,\Lambda_3\rangle.
\lb{ns57}
\ee
Before we reduce this expression further, we illustrate the formula for
the BK solution of $G=SU(2)$. Then we have
\be
\Lambda_1=w\tau_1,\qquad\Lambda_2=w\tau_2,\qquad\Lambda_3=\tau_3
\lb{ns58}
\ee
with the boundary conditions
\be
w(0)=1,\qquad w(\infty)=-1.
\lb{ns59}
\ee
This implies %
$\Lambda_{(0)}=1$, %
$\Lambda_{(\infty)}=\mbox{Ad}(\exp(\pi\tau_3))\Lambda_{(0)}$, %
i.e.\ $Z=-\pi\tau_3$ and thus %
$\langle Z,\Lambda_3\rangle=-\pi\langle \tau_3,\tau_3\rangle$. %
If we choose the scalar product as
\[\langle X,Y\rangle =-1/(2\pi)^2tr(XY)\]
we obtain thus
\be\int Q(\tilde A)=\frac{1}{2}.
\lb{ns60}
\ee
The scalar product is chosen such that the second term in (\ref{ns46}),
which is proportional to the winding number of $g$, is an integer. We
have
\be
\int_{S^3} tr(\theta\wedge\theta\wedge\theta)=-24\pi^2\,\mbox{deg}(g).
\ee
The result (\ref{ns60}) was obtained before by Moss and Wray
\cite{moss}.
\subsection{Further reductions}
Let us first express the right hand side of (\ref{ns57}) in terms of
the classifying element $H_\lambda$ in (\ref{ns9}). Since %
$H_\lambda := L\lambda(2\pi i\sigma_3)=-4\pi
L\lambda(\tau_3)=-4\pi\Lambda_3$ %
we obtain from (\ref{ns57})
\be
\int Q(\tilde A) = \langle Z,H_\lambda\rangle.
\lb{ns61}
\ee

In the course of the evaluation of $\langle Z,H_\lambda\rangle$ we make
now use of various tools which are developed in Appendix A. There we
show that $S(\lambda)$ is a basis of a root system $R(\lambda)$
(Theorem A1) and that
\be
H_\lambda = 2{\varrho(\lambda)}^\ast
	 := \sum_{\alpha\in {R(\lambda)}_+}\alpha^\ast,
\lb{ns62}
\ee
where $\alpha^\ast$ denotes the inverse root corresponding to $\alpha$
(see eq. (\ref{nsa16})). If the root system $R(\lambda)$ is reducible
with the decomposition
\be
R(\lambda)=\bigcup_j R_j(\lambda)
\lb{ns63}
\ee
into mutually orthogonal irreducible root systems $R_j(\lambda)$ then
$\langle Z,H_\lambda\rangle$ splits into the sum
\be
\langle Z,H_\lambda\rangle = \sum_j\langle
Z,2{\varrho_j(\lambda)}^\ast\rangle,
\qquad
2{\varrho_j(\lambda)}^\ast = \sum_{\alpha\in
{R_j(\lambda)}_+}\alpha^\ast.
\lb{ns64}
\ee
It is clear that $S(\lambda)$ decomposes correspondingly:
\be
S(\lambda)=\bigcup_j S_j(\lambda),\qquad
S_j(\lambda)=S(\lambda)\cap R_j(\lambda).
\lb{ns65}
\ee
It suffices, therefore, to compute $\langle Z,H_\lambda\rangle$ for
irreducible root systems which are classified by the well-known Dynkin
diagrams.

Thus from now on $R(\lambda)$ is assumed to be irreducible. For the
computation of $\langle Z,H_\lambda\rangle$ we expand
$2{\varrho(\lambda)}^\ast$ with respect to the basis $S(\lambda)$ as in
(\ref{nsa8})
\be
2{\varrho(\lambda)}^\ast
=
\sum_{\alpha\in {S(\lambda)}}
n_\alpha\,
\frac{2\alpha}{\langle\alpha_L,\alpha_L\rangle},
\lb{ns66}
\ee
and $Z$ is expanded as in (\ref{nsa17}) with respect to the dual basis
$\{h_\alpha\}_{\alpha\in {S(\lambda)}}$ of $S(\lambda)$
\be
Z=\sum_{\alpha\in {S(\lambda)}} Z^\alpha h_\alpha + Z_\perp,\qquad
Z_\perp\in{\langle S(\lambda)\rangle}^\perp.
\lb{ns67}
\ee
In (\ref{ns66}) $\langle\alpha_L,\alpha_L\rangle$ is the square of the
length of a long root, and one knows that $n_\alpha$ are all positive
integer numbers which are determined by the Cartan matrix $N(\lambda)$
of $R(\lambda)$ (see eq.\ (\ref{nsa9})). The coefficients $Z^\alpha$ in
(\ref{ns67}) have to be chosen such that
\be
\exp(2\pi i Z^\alpha)= \frac{w^\alpha (\infty)}{w^\alpha (0)}
\lb{ns68}
\ee
(see (\ref{nsa18})), where the amplitudes $w^\alpha$ are defined by the
expansion of $\Lambda_+$,
\be
\Lambda_+=\sum_{\alpha\in {S(\lambda)}} w^\alpha e_\alpha
\lb{ns69}
\ee
in terms of base elements $e_\alpha$ of the root spaces $L_\alpha$.

Collecting all this we find
\begin{eqnarray}
&&\nonumber\\
\langle Z,H_\lambda\rangle
& = &
\frac{2}{\langle\alpha_L,\alpha_L\rangle}
\sum_{\alpha\in {S(\lambda)}} Z^\alpha n_\alpha,
\lb{ns70}\\
&&\nonumber\\
Z^\alpha
& = &
\frac{1}{2\pi i}\ln\frac{w^\alpha (\infty)}{w^\alpha (0)}\quad\in{\cal
R},
\lb{ns71}\\
&&\nonumber\\
n_\alpha
& = &
2\sum_{\beta\in {S(\lambda)}} (N(\lambda)^{-1})_{\alpha\beta}
\frac{\langle\alpha_L,\alpha_L\rangle}{\langle\beta,\beta\rangle}
\quad \in {\cal N}.\lb{ns72}\\
&&\nonumber
\end{eqnarray}
The coefficients $n_\alpha$ are listed in Table \ref{taba1}, \ref{taba2}
of Appendix A for all simple root systems.

Clearly, the coefficients $Z^\alpha=\alpha(Z)$ in (\ref{ns71}) are only
determined up to integer numbers. A corresponding change $Z\mapsto Z +
\Delta Z$ induces an additive contribution in (\ref{ns70}) of the form
\be
\langle \Delta Z,H_\lambda\rangle
= \frac{2}{\langle\alpha_L,\alpha_L\rangle}\, k(\lambda)\,m,\qquad
m\in{\cal Z},
\lb{ns73}
\ee
where $k(\lambda)$ is a positive integer which is determined by the
root system $R(\lambda)$ and which we have listed in Table 1 for the
simple root systems.
\begin{table}
\[
\begin{array}{rlcrll}
A_l: &
k  =  \left \{
\begin{array}{ll}
2    & \mbox{ for }l=2m,\\
1    & \mbox{ otherwise},
\end{array} \right.
& \quad\! &
B_l: &
k  =  2,
\\
&&&&&
\\
C_l: &
k  =  \left \{
\begin{array}{ll}
2    & \mbox{ for }l=2m,\\
1    & \mbox{ otherwise},
\end{array} \right.
& &
D_l: &
k  =  \left \{
\begin{array}{ll}
2    & \mbox{ for }l=4m,4m+1,\\
1    & \mbox{ otherwise},
\end{array} \right.
\\
&&&&&
\\
E_l: &
k  =  \left \{
\begin{array}{ll}
2    & \mbox{ for }l=6,8,\\
1    & \mbox{ for }l=7,
\end{array} \right.
& &
F_4: &
k  =   2,
\\
&&&&&
\\
G_2: &
k  =   2.
&  &
&&
\end{array}
\]
\caption{The factors $k(\lambda)$ for the simple root systems}
\end{table}

We show now that
\be
Z^\alpha=\alpha(Z) \in \frac{1}{2} {\cal Z}.
\lb{ns74}
\ee
This relies on the consequence (\ref{ns51}) of the YM equations,
according to which the phases of all $w^\alpha$, $\alpha\in
S(\lambda)$, are constant. Hence (\ref{ns68}) reduces to
\[\exp(2\pi i Z^\alpha)=\pm 1,\]
i.e.
\[Z^\alpha\in\frac{1}{2}{\cal Z}.\]

We can now use this fact in (\ref{ns70}) and find, using also
(\ref{ns73}),
\be
\langle Z,H_\lambda\rangle
\in
\frac{2}{\langle\alpha_L,\alpha_L\rangle}\,
\frac{k(\lambda)}{2}\,{\cal Z}.
\lb{ns75}
\ee

Let us work this out for $G=SU(n+1)$ and $R(\lambda)=A_n$, with the
scalar product
\be
\langle X,Y \rangle=-\frac{1}{(2\pi)^2}\, tr(XY).
\lb{ns76}
\ee
Then all roots have length $\sqrt{2}$. According to Table 1
$k(\lambda)=1$ for $n$ odd, and $k(\lambda)=2$ for $n$ even. Thus
\be
\langle Z,H_\lambda\rangle
\in
\left \{
\begin{array}{ll}
	 \frac{1}{2} {\cal Z} & \mbox{ for }n\mbox{ odd},\\
\phantom{\frac{1}{2}}{\cal Z} & \mbox{ for }n\mbox{ even}.
\end{array}
\right.
\lb{ns77}
\ee

The definition of $Q(\tilde A)$ depends on the normaliszation of the
inner product, and hence also the combined result (\ref{ns61}) and our
various formulae for $\langle Z,H_\lambda\rangle$:
\begin{eqnarray}
&&\nonumber\\
\int Q(\tilde A)
& = &
\sum_{\mbox{irr.}}\,\langle Z,2{\varrho (\lambda)}^\ast\rangle
\lb{ns78}\\
&&\nonumber\\
& = &
\sum_{\mbox{irr.}}
\frac{2}{\langle\alpha_L,\alpha_L\rangle}
\sum_{\alpha\in {S(\lambda)}}
\alpha (Z)\,n_\alpha
\lb{ns79} \\
&&\nonumber\\
& \in &
\sum_{\mbox{irr.}}
\frac{2}{\langle\alpha_L,\alpha_L\rangle}
\frac{k(\lambda)}{2}{\cal Z}.\nonumber\\
&&\nonumber
\end{eqnarray}
($\sum_{\mbox{irr.}}$ indicates the sum over the irreducible components
in the decomposition (\ref{ns63}) of the root system $R(\lambda)$.)

For a simple gauge group and a given normalisation of the scalar
product one will define the Chern-Simons number as
\be
N_{CS}:=\#\int Q(\tilde A),
\ee
where the prefactor is chosen such that the changes $\Delta N_{CS}$ of
$N_{CS}$ under global gauge transformations can be any integer. This
was illustrated for $SU(2)$ at the end of section 6.1 (see
eq.\ (\ref{ns60})). The normalization factor $1$ in (\ref{ns60})
remains the same for all $SU(n)$, if the scalar product is chosen
according to (\ref{ns76}). We do not discuss here the normalization
question for the other compact Lie groups.
\section{Concluding remarks}
In concluding, we emphasize, that most of the results of the present
paper do not rely on the detailed form of the coupled field equations,
but are mainly based on bundle and group theoretical considerations.
This is in particular the case for our classification of the possible
regular, spherically symmetric solutions of the EYM equations (sect.\ 2
and Appendix B). We have started a detailed study of the field
equations which we have brought into an explicit and well-adapted
form.
We expect that the set of $A_1$-vectors in the open Weyl chamber (see
Appendix B), for which there exist regular solutions, is further
constraint because one is dealing with a singular boundary value
problem. This does, however, not seem to be the case for the gauge
groups $SU(n)$, as we learned recently from H.P.\ K\"unzle
\cite{kunzle}, who has been able to construct numerically new regular
solutions for $SU(3)$.

\bigskip
\noindent{\large Acknowledgments}
\medskip

We would like to thank A.\ Wipf for useful discussions on the possible
role of EYM solitons in baryon violating processes. We are also
grateful to H.P.\ K\"unzle for informative discussions and for showing
us his recent numerical results of new EYM solutions. This work was
supported in part by the Swiss National Science Foundation.
\vskip 2.3\baselineskip

\appendix
\noindent{\Large APPENDIX}
\vskip -\baselineskip

\vskip -0\baselineskip

\section{Mathematical tools and detailed proofs}
\setcounter{figure}{0}
\renewcommand{\thefigure}{\Alph{section}\arabic{figure}}
\setcounter{table}{0}
\renewcommand{\thetable}{\Alph{section}\arabic{table}}
\setcounter{equation}{0}
\renewcommand{\theequation}{\Alph{section}\arabic{equation}}
\setcounter{definition}{0}
\renewcommand{\thedefinition}{\Alph{section}\arabic{definition}}
\setcounter{theorem}{0}
\renewcommand{\thetheorem}{\Alph{section}\arabic{theorem}}
\setcounter{proposition}{0}
\renewcommand{\theproposition}{\Alph{section}\arabic{proposition}}
\setcounter{corollary}{0}
\renewcommand{\thecorollary}{\Alph{section}\arabic{corollary}}
\setcounter{lemma}{0}
\renewcommand{\thelemma}{\Alph{section}\arabic{lemma}}
In this Appendix we prove some mathematical facts which are used in the
main body of the paper.

For the formulation of the next theorem we use the following notation
which was also adopted in the main text.

We consider a compact connected Lie group $G$ with a fixed maximal
torus $T$ and the correspondig integral lattice $I$. $R$ denotes the
set of real roots and $S$ a basis of $R$ with corresponding fundamental
(open) Weyl chamber $K(S)$.

For the evaluation of the Chern-Simons number in sect.\ 6 we need the
following
\begin{theorem} Let $H\in K(S)$ and
\be
S(H)=\{\, \alpha\in R \mid \alpha (H)=2\,\}\;\subset R_+.
\lb{nsa1}
\ee
If $S(H)$ is not empty then $S(H)$ is the basis of a root system
$R(H)\subset R$.
\end{theorem}
We shall show that $S(H)$ is an abstract fundamental system in the
sense of the following definition, and then use known results from the
mathematical literature.
\begin{definition} An {\em abstract fundamental system} is a non-empty,
finite, linearly independent subset $F=\{\alpha_1,\ldots,\alpha_l\}$ of
a Euclidean space such that for any $\alpha_i,\alpha_j\in F$ with
$\alpha_i\neq\alpha_j$ the value %
$2\langle \alpha_i,\alpha_j\rangle/
  \langle \alpha_j,\alpha_j\rangle$ %
is a non-positive integer.
\end{definition}
\noindent{\sc Proof of Theorem A1}{\hspace{0.75em}}First, we establish
the following three properties of $S(H)$:
\begin{list}{(\roman{c_one})}
	    {\usecounter{c_one}}
\item If $\alpha,\beta\in S(H)$ then $\alpha-\beta$ is not a root.
\item For $\alpha,\beta\in S(H)$ with  $\alpha\neq\beta$, we have
$\langle \alpha,\beta\rangle\leq 0$.
\item The set $S(H)$ is linearly independent.
\end{list}

The first statement is obvious. If $\alpha-\beta$ would be a root it
would have to vanish on $H$, which is , however, impossile because $H$
is in the open Weyl chamber $K(S)$ on which all positive roots have
strictly positive values.

Turning to (ii), suppose that $\langle \alpha,\beta\rangle > 0$ for a
pair  $\alpha,\beta\in S(H)$. Then the Cartan number $n_{\alpha\beta}$
for the roots $\alpha,\beta\in S(H)$ would also be strictly positive
and thus also  $n_{\beta\alpha}$. Therefore, the product
$n_{\alpha\beta}n_{\beta\alpha}$ must have one of the values $1,2,3$
($4$ is excluded for $\alpha\neq\beta$). This leaves two
possibilities:
either $n_{\alpha\beta}=1$ or $n_{\beta\alpha}=1$. In the second case
we have %
$\alpha -\beta =\alpha-n_{\beta\alpha}\beta
	       =\alpha-\langle \beta^\ast,\alpha\rangle\beta
	       =s_\beta\alpha\in R$, %
where $s_\beta$ is the Weyl reflection belonging to $\beta$, and this
is impossible according to (i). Similarly, the first possibility is
also excluded.

Having established $\langle \alpha,\beta\rangle\leq 0$ for
$\alpha,\beta\in S(H)$, we can prove the linear independence of $S(H)$
by a standard argument. A linear dependence of $S(H)$ would imply an
equation of the form
\be
 \sum_{\beta}  m_\beta  \beta
=\sum_{\gamma} n_\gamma \gamma,\qquad m_\beta,n_\gamma \geq 0,
\nonumber
\ee
where $\beta$ and $\gamma$ run through disjoint subsets of $S(H)$.
Denoting one side of this equation by $\phi$, we have
\be
{|\phi|}^2=
\sum_{\beta,\gamma}m_\beta n_\gamma\,\langle\beta,\gamma\rangle \leq 0
\nonumber
\ee
and thus $\phi=0$. But then we have
\be
0=\langle\phi,H\rangle=\sum_\beta  m_\beta \langle\beta ,H\rangle
		      =\sum_\gamma n_\gamma\langle\gamma,H\rangle.
\nonumber
\ee
Since $H\in K(S)$ and $S(H)\subset R_+$ this is only possible for
$m_\beta=n_\gamma=0$ for all $\beta,\gamma$.

{}From what has been shown so far it is clear that $S(H)$ is an
abstract
fundamental system in the sense of the definition given above. To such
a system $F$ we can consider the Weyl group $W$ generated by all
reflections. One can prove quite easily that $W\cdot F$ is a root
system. In addidion, it turns out that $F$ is also a fundamental system
of roots of $W\cdot F$. Proofs of these facts can be found in
sect.\ 2.12 of \cite{samelson}. Another possibility to establish these
statements is to construct all possible abstract fundamental systems
and to verify them in each case: The simple (indecomposable) such
systems are described Dynkin diagrams. (To each of these belongs an
irreducible root system of which $F$ is basis.)

Before we draw a useful corollarry to Theorem A1 we note that the
decomposition of $LT$ into the linear span $\langle S(H)\rangle$ of
$S(H)$ and its orthogonal complement
\be
LT=\langle S(H)\rangle\oplus {\langle S(H)\rangle}^\perp
\lb{nsa2}
\ee
is independent of the Ad$G$ -invariant metric. Indeed, %
$\langle S(H)\rangle$ %
is also equal to the linear span of the set of inverse roots %
$S(H)^\ast:=\{\,\alpha^\ast\mid\alpha\in S(H)\,\}$ and %
${\langle S(H)\rangle}^\perp=\bigcap_{\alpha\in
S(H)}\mbox{ker}\,\alpha$%
. (We recall that the inverse roots $\alpha^\ast\in LT$ are independent
of the choice of the inner product; see, e.g., chapt.\ V of
Ref.\ \cite{brocker}.)

In sect. 6.2 we have used the following
\begin{corollary} With the same notation as in theorem A1 we have for
$H_{\parallel}$ in the decomposition $H=H_{\parallel}+H_{\perp}$ with
respect to
(\ref{nsa2})
\be
H_{\parallel}= 2{\varrho (H)}^\ast
	    := \sum_{\alpha\in R(H)_+} \alpha^\ast\,
	    \in I.
\lb{nsa3}
\ee
\end{corollary}
\noindent{\sc Proof}{\hspace{0.75em}}{}From Theorem A1 it follows that
$S(H)^\ast$ is a basis of the root system $R(H)^\ast$ and therefore the
sum of positive inverse roots, $2\varrho(H)^\ast$, satisfies (see
sect.\ V.4 in \cite{brocker})
\be
\langle 2\varrho(H)^\ast,\alpha\rangle=2 \qquad \mbox{ for all }
\alpha\in S(H).
\lb{nsa4}
\ee
On the other hand, by (\ref{nsa1}) we have $\alpha (H_{\parallel}) =
\alpha (H) = 2$ for all $\alpha \in S(H)$. This proofs (\ref{nsa3}).

The quantity $2\varrho (H)^\ast$ plays an inportant role in our
evaluation of the Chern-Simons numbers and we need a more explicit
expression for it.

First of all we note that if the root system $R(H)$ is reducible with
the decomposition into mutually orthogonal irreducible root systems
\be
R(H) = \bigcup_j R_j(H),
\lb{nsa5}
\ee
then $S(H)$ and $\varrho (H)^\ast$ split correspondingly:
\be
\begin{array}{rcl}
 S(H)             & = & \bigcup_j S_j(H),\\
		  &   & \\
2\varrho (H)^\ast & = & \sum_j 2\varrho_j (H)^\ast,
\end{array}
\quad
\begin{array}{rcl}
S_j(H)              &    =    & S(H)\cap R_j(H),\\
		    &         & \\
2\varrho_j (H)^\ast &    =    & \sum_{\alpha \in R_j(H)_+}
\alpha^\ast.
\end{array}
\lb{nsa6}
\ee

We expand now $2\varrho_j (H)^\ast$ in terms of the basis $S(H)_j$
\be
2\varrho_j (H)^\ast = \sum_{\alpha \in S_j(H)} n_\alpha\,
\frac{2\alpha}{\langle
\alpha_L,\alpha_L\rangle} .
\lb{nsa8}
\ee
Here, ${\langle \alpha_L,\alpha_L\rangle}$ is the square of the length
of a long root. (We recall that at most two root lengths occur in an
irreducible root system.) Since $\alpha^\ast =
2\alpha/\langle\alpha,\alpha\rangle$ it is clear that the coefficients
$n_\alpha$ in (\ref{nsa8}) are positive integers. In addition, they are
independent of the choice of the scalar product. An explicit formula in
therms of the Cartan matrix $N_j(H)$ of $S_j(H)$ follows by writing out
the equation $\langle 2\varrho_j(H)^\ast, \beta\rangle = 2$, for every
$\beta \in S_j(H)$. One finds
\be
n_\alpha = 2\sum_{\beta \in S_j(H)} ({N_j(H)}^{-1})_{\alpha \beta}\,
\frac{\langle \alpha_L,\alpha_L\rangle}
     {\langle\beta,\beta\rangle} .
\lb{nsa9}
\ee

In Table \ref{taba1}, \ref{taba2} we have listed the expansion
coefficients $n_\alpha$ for all irreducible (simple) root systems.
\def\sdinkfill{$\mathord-
\mkern-6mu\cleaders\hbox{$\mkern-2mu\mathord-\mkern-2mu$}
\hfill
\mkern-6mu
\mathord-$}

\def\sdink{%
\hbox to 1.55cm {\sdinkfill}}

\def\sdinketc{\hbox to 2.5cm {\sdinkfill$\;\cdots\;$\sdinkfill}}

\def\ddinkfill{$\mathord=
\mkern-6mu\cleaders\hbox{$\mkern-2mu\mathord=\mkern-2mu$}
\hfill
\mkern-6mu
\mathord=$}

\def\rddink{%
\hbox{\rlap{\hbox to 1.55cm {\hfill$\mathord >$\hfill}}%
\hbox to 1.55cm {\ddinkfill}}}

\def\lddink{%
\hbox{\rlap{\hbox to 1.55cm {\hfill$\mathord <$\hfill}}%
\hbox to 1.55cm {\ddinkfill}}}

\def\tdinkfill{$\mathord\equiv
\mkern-6mu\cleaders\hbox{$\mkern-2mu\mathord\equiv\mkern-2mu$}
\hfill
\mkern-6mu
\mathord\equiv$}

\def\rtdink{%
\hbox{\rlap{\hbox to 1.55cm {\hfill$\mathord >$\hfill}}%
\hbox to 1.55cm {\tdinkfill}}}

\def\dcirc#1{{\displaystyle\mathop{\bigcirc}%
_{\hbox{\vphantom {\big (}{\hbox to 0pt{\hss $\scriptstyle
#1$\hss}}}}}}

\def\ucirc#1{{\displaystyle\mathop{\bigcirc}%
^{\hbox{\vphantom {\big (}{\hbox to 0pt{\hss $\scriptstyle
#1$\hss}}}}}}

\def\A_l{%
{\renewcommand{\arraycolsep}{-1pt}
\begin{array}{lccccccccccc}
\hbox to 2em{\hfill $A_l\colon$}\qquad&
\ucirc{1\cdot l}&
\sdink&
\ucirc{2\cdot (l-1)}&
\sdinketc&
\ucirc{m\cdot (l-m+1)}&
\sdinketc&
\ucirc{(l-2)\cdot 3}&
\sdink&
\ucirc{(l-1)\cdot 2}&
\sdink&
\ucirc{l\cdot 1},
\end{array}}
}%

\def\B_l{%
{\renewcommand{\arraycolsep}{-1pt}
\begin{array}{lccccccccccc}
\hbox to 2em{\hfill $B_l\colon$}\qquad&
\ucirc{1\cdot 2l}&
\sdink&
\ucirc{2\cdot (2l-1)}&
\sdinketc&
\ucirc{m\cdot (2l-m+1)}&
\sdinketc&
\ucirc{(l-2)\cdot (l+3)}&
\sdink&
\ucirc{(l-1)\cdot (l+2)}&
\rddink&
\ucirc{l\cdot (l+1)},
\end{array}}
}%

\def\C_l{%
{\renewcommand{\arraycolsep}{-1pt}
\begin{array}{lccccccccccc}
\hbox to 2em{\hfill $C_l\colon$}\qquad&
\ucirc{2\cdot (2l-1)}&
\sdink&
\ucirc{4\cdot (2l-2)}&
\sdinketc&
\ucirc{2m\cdot (2l-m)}&
\sdinketc&
\ucirc{2(l-2)\cdot(l+2)}&
\sdink&
\ucirc{2(l-1)\cdot(l+1)}&
\lddink&
\ucirc{l^2},
\end{array}}
}%

\def\D_l{%
{\renewcommand{\arraycolsep}{-1pt}
 \renewcommand{\arraystretch}{.7}
\begin{array}{lccccccccccc}
\hbox to 2em{\hfill $D_l\colon$}\qquad&
\ucirc{1\cdot (2l-2)}&
\sdink&
\ucirc{2\cdot (2l-3)}&
\sdinketc&
\ucirc{m\cdot (2l-m-1)}&
\sdinketc&
\ucirc{(l-3)\cdot (l+2)}&
\sdink&
\ucirc{(l-2)\cdot (l+1)}&
\sdink&
\ucirc{l(l-1)/2},\\
&&&&&&&&&\mathord{\Big  |}&&\\
&&&&&&&&&\mathord{\bigg |}&&\\
&&&&&&&&&\dcirc{l(l-1)/2} &&\\
\end{array}}
}%

\def\Esix{%
{\renewcommand{\arraycolsep}{-1pt}
 \renewcommand{\arraystretch}{.7}
\begin{array}{lccccccccc}
\hbox to 2em{\hfill $E_6\colon$}\qquad&
\ucirc{16}&
\sdink&
\ucirc{30}&
\sdink&
\ucirc{42}&
\sdink&
\ucirc{30}&
\sdink&
\ucirc{16},\\
&&&&&\mathord{\Big  |}&&&&\\
&&&&&\mathord{\bigg |}&&&&\\
&&&&&\dcirc{22} &&&&\\
\end{array}}
}%

\def\Eseven{%
{\renewcommand{\arraycolsep}{-1pt}
 \renewcommand{\arraystretch}{.7}
\begin{array}{lccccccccccc}
\hbox to 2em{\hfill $E_7\colon$}\qquad&
\ucirc{34}&
\sdink&
\ucirc{66}&
\sdink&
\ucirc{96}&
\sdink&
\ucirc{75}&
\sdink&
\ucirc{52}&
\sdink&
\ucirc{27},\\
&&&&&\mathord{\Big  |}&&&&&&\\
&&&&&\mathord{\bigg |}&&&&&&\\
&&&&&\dcirc{49} &&&&&&\\
\end{array}}
}%

\def\Eeight{%
{\renewcommand{\arraycolsep}{-1pt}
 \renewcommand{\arraystretch}{.7}
\begin{array}{lccccccccccccc}
\hbox to 2em{\hfill $E_8\colon$}\qquad&
\ucirc{92}&
\sdink&
\ucirc{182}&
\sdink&
\ucirc{270}&
\sdink&
\ucirc{220}&
\sdink&
\ucirc{168}&
\sdink&
\ucirc{114}
\sdink&
\ucirc{58}.\\
&&&&&\mathord{\Big  |}&&&&&&&&\\
&&&&&\mathord{\bigg |}&&&&&&&&\\
&&&&&\dcirc{136} &&&&&&&&\\
\end{array}}
}%

\def\Ffour{%
{\renewcommand{\arraycolsep}{-1pt}
 \renewcommand{\arraystretch}{.7}
\begin{array}{lccccccc}
\hbox to 2em{\hfill $F_4\colon$}\qquad&
\ucirc{22}&
\sdink&
\ucirc{42}&
\rddink&
\ucirc{60}&
\sdink&
\ucirc{32},\\
\end{array}}
}%

\def\Gtwo{%
{\renewcommand{\arraycolsep}{-1pt}
 \renewcommand{\arraystretch}{.7}
\begin{array}{lccc}
\hbox to 2em{\hfill $G_2\colon$}\qquad&
\ucirc{10}&
\rtdink&
\ucirc{18},\\
\end{array}}
}
\begin{table}
\[
\begin{array}{l}
\A_l\\
\hbox{\vphantom{\bigg)}}\\
\B_l\\
\hbox{\vphantom{\bigg)}}\\
\C_l\\
\hbox{\vphantom{\bigg)}}\\
\D_l\\
\end{array}
\]
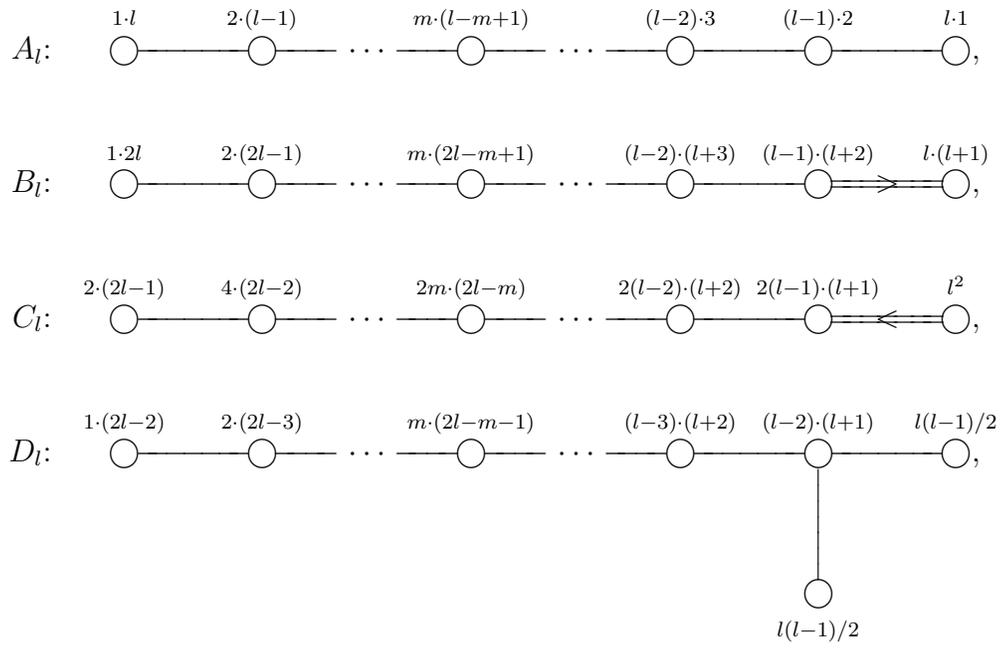
\caption{Expansion coefficients $n_\alpha$ in (\protect{\ref{nsa8}})
of the sum of positive inverse roots for the classical root
systems.\lb{taba1}}
\end{table}
\begin{table}
\[
\begin{array}{l}
\Gtwo\\
\hbox{\vphantom{\bigg)}}\\
\Ffour\\
\hbox{\vphantom{\bigg)}}\\
\Esix\\
\Eseven\\
\Eeight\\
\end{array}
\]
\caption{Expansion coefficients $n_\alpha$ in (\protect{\ref{nsa8}})
of the sum of positive inverse roots for the exceptional root
systems.\lb{taba2}}
\end{table}
Another tool which is used in sect.\ 5 is
\begin{proposition} Let $\Lambda, \tilde \Lambda$ be two homomorphisms
from $LSU(2)$ to $LG$ with $H:=\Lambda (2\pi i \sigma_3) = \tilde
\Lambda (2\pi i \sigma_3) \in K(S)$, then these are conjugated by an
element in $T$, i.e. there exists an element $Z \in LT$ such that
\be
\tilde \Lambda = \mbox{\rm Ad}(\exp Z) \circ \Lambda.
\lb{nsa10}
\ee
\end{proposition}
\noindent{\sc Proof}{\hspace{0.75em}}We use the following standard
basis of $LSU(2)_{\cal C}$:
\be
2\pi i\sigma_3=2\pi i\left( \begin{array}{cr}1 & 0 \\ 0&-1 \end{array}
\right),
\qquad
E_+=\left( \begin{array}{cc}0 & 1 \\ 0&0 \end{array} \right),
\qquad
E_-=\left( \begin{array}{rc}0 & 0 \\ -1&0 \end{array} \right).
\lb{nsa11}
\ee
Since $\Lambda$ is a homomorphism we have with the notation
$\Lambda_\pm :=\Lambda (E_\pm)$
\be
\frac{1}{2\pi i}[H,\Lambda_\pm]=\pm 2\,\Lambda_\pm, \qquad
	  [\Lambda_+,\Lambda_-]=-\frac{1}{2\pi i}H.
\lb{nsa12}
\ee
{}From the first two equations and the assumption about $H$ we conclude
as in sect.\ 3 that $\Lambda_+$ is in the direct sum of root spaces
$L_\alpha$ for $\alpha\in S(H)$. (Note, that $S(H)$ is not empty since
$\Lambda_\pm=0$ would imply $H=0\notin K(S)$.) Hence we can expand
$\Lambda_+$ relative to a basis $\{e_\alpha\}_{\alpha\in S(H)}$
\be
\Lambda_+=\sum_{\alpha\in S(H)} w^\alpha e_\alpha.
\lb{nsa13}
\ee
If we insert this and $\Lambda_-=c(\Lambda_+)$ ($c$ = conjugation in
$LG_{\cal C}$) in the third relation of (\ref{nsa12}) and use that
$S(H)$ is a simple system of roots (Theorem A1), then we obtain with
the well-known orthogonality and commutation relations
\be
\sum_{\alpha\in S(H)} {|w^\alpha |}^2 \langle
e_\alpha,c(e_\alpha)\rangle\,
2\pi i\,\alpha
=-\frac{1}{2\pi i}H,
\lb{nsa14}
\ee
The parallel compontent relative to (\ref{nsa2}) gives with
(\ref{nsa3}) and (\ref{nsa8})
\be
{|w^\alpha |}^2 = \frac{1}{{(2\pi)}^2}\,
\frac{1}{\langle e_\alpha,c(e_\alpha)\rangle}\,
\frac{2n_\alpha}{\langle\alpha_L,\alpha_L\rangle}\;
>0
\lb{nsa15}
\ee
and $H_{\perp}$ must vanish. (\ref{nsa15}) shows that the amplitudes
$w^\alpha$ are unique up to a phase and that none of them vanishes. In
addition we have with (\ref{nsa3}) also
\be
H=2\varrho (H)^\ast\,\in I ,\qquad H=\Lambda(2\pi i\sigma_3).
\lb{nsa16}
\ee

Now we can construct the element $Z$ in (\ref{nsa10}). It is convenient
to introduce the basis $\{{h_\alpha}\}_{\alpha\in S(H)}$ which is dual
to the base $S(H)$. We claim that
\be
Z:=\sum_{\alpha\in S(H)}Z^\alpha h_\alpha + Z_{\perp},
\lb{nsa17}
\ee
with $Z$ chosen such that
\be
\exp(2\pi iZ^\alpha) = \frac{\tilde w^\alpha}{w^\alpha}
\lb{nsa18}
\ee
and $Z_{\perp}\in {\langle S(H)\rangle}^{\perp}$, satisfies
(\ref{nsa10}). Indeed, (\ref{nsa10}) is obviously correct for the base
element $2\pi i\sigma_3$. Furthermore,
\begin{eqnarray}
&&\nonumber\\
\mbox{Ad}(\exp Z)\Lambda_+
& = & \sum_{\alpha\in S(H)} w^\alpha\mbox{Ad}(\exp Z)e_\alpha
\nonumber\\
&&\nonumber\\
& = & \sum_{\alpha\in S(H)} w^\alpha\exp (2\pi i \alpha (Z)) e_\alpha
\nonumber\\
&&\nonumber\\
& = & \sum_{\alpha\in S(H)} w^\alpha\exp (2\pi i Z^\alpha)   e_\alpha
\nonumber\\
& = & \tilde\Lambda_+.
\nonumber\\
&&\nonumber
\end{eqnarray}

At this point it is simple to prove the following lemma, which is used
at the end of sect. 4.
\begin{lemma} Let $\Lambda\colon LSU(2)\rightarrow LG$ be a
homomorphism and suppose that $H:=\Lambda (2\pi i\sigma_3)\in K(S)$. If
an element $A\in LT$ satisfies $[A,\Lambda_+]=0$ then $A\in {\langle
S(H)\rangle}^{\perp}$.
\end{lemma}
\noindent{\sc Proof}{\hspace{0.75em}}If we insert (\ref{nsa13}) in
$[A,\Lambda_+]=0$ we obtain
\be
0=\sum_{\alpha\in S(H)} w^\alpha [A,e_\alpha]
 =\sum_{\alpha\in S(H)} w^\alpha \alpha (A)e_\alpha.
\nonumber
\ee
Since all the $w^\alpha,\alpha\in S(H)$, are different from zero, this
implies $\alpha (A)=0$ for all $\alpha\in S(H)$, i.e. $A\in
\bigcap_{\alpha\in S(H)}\mbox{ker}\,\alpha={\langle
S(H)\rangle}^{\perp}$.
\section{The classifying homomorphisms for EYM solitons}
\setcounter{figure}{0}
\setcounter{table}{0}
\setcounter{equation}{0}
\setcounter{definition}{0}
\setcounter{theorem}{0}
\setcounter{proposition}{0}
\setcounter{corollary}{0}
\setcounter{lemma}{0}
At the end of sect.\ 2 we noticed that the classifying homomorphism
$\lambda\colon U(1)\rightarrow G$ of spherically symmetric principal
$G$-bundles, satisfying $L\lambda(2\pi i)\in I\cap\overline{K(S)}$, is
further restricted for regular EYM solutions: $\lambda$ must have an
extension $\tilde\lambda\colon K\rightarrow G$, with $K=SO(3)$ or
$SU(2)$.

There is a systematic procedure to determine all lattice ponts in
$I\cap\overline{K(S)}$ for wich this is possible. A sketch of this is
given below, which works for any gauge group $G$ if we restrict ourself
to the generic case of points in the open fundamental Weyl chamber
$K(S)$.

It is useful to introduce the following terminology: A vector $H\in LT$
is an {\em $A_1$-vector} if there exist two elements $\Lambda_\pm\in
LG_{\cal C}$ such that $H$, $\Lambda_\pm$ satisfy the $A_1$-commutation
relations (\ref{nsa12}). (For instance, any inverse root of a
semisimple Lie algebra is an $A_1$-vector.) {}From what has been said
we
have to determine all $A_1$-vectors in $K(S)$. Below we show that the
enumeration of this set can be reduced to quite a different problem
which has been solved completely long ago by Dynkin \cite{dynkin}. In
his classification of so-called regular subalgebras of semisimple Lie
algebras, Dynkin had to describe certain subsets of the root system
which he called $\Pi$-systems. These are defined as follows:
\begin{definition} A subset $\Sigma$ of a root system $R$ is a {\em
$\Pi$-system} if:
\begin{list}{(\roman{c_one})}
	    {\usecounter{c_one}}
\item $\alpha,\beta\in \Sigma$ implies that $\alpha-\beta$ is not a
root;
\item the set $\Sigma$ is linearly independent.
\end{list}
\end{definition}
It is important to note that the proof of Theorem A1 implies that any
$\Pi$-system $\Sigma$ is a basis of a root system $R(\Sigma)\subset
R$.

Dynkin has given an explicit description of all $\Pi$-systems of the
root system of a semisimple Lie algebra which we do not want to repeat
here. We shall show, however, that the knowledge of the $\Pi$-systems
also enables one to determine all $A_1$-vectors in $K(S)$. In order to
avoid unimportant complications we assume that the gauge group $G$ is
semisimple.

As a useful tool we consider the map $\kappa$, which associates to any
$\Pi$-system $\Sigma$ the following element in the ${\cal Z}$-subspace
${\rm Span}_{\cal Z}(R^\ast)$, spanned by the set $R^\ast$ of inverse
roots:
\be
\kappa(\Sigma):=\sum_{\alpha\in R(\Sigma)_+} \alpha^\ast.
\lb{nsb1}
\ee
This map respects the natural actions of the Weyl group $W$,
\be
\kappa(w\cdot\Sigma)=w\cdot\kappa(\Sigma)\qquad \mbox{for all $w\in
W$},
\lb{nsb2}
\ee
and induces thus a map between the corresponding orbit spaces. Now we
use the fact that the orbit of an arbitrary point in the ${\cal
R}$-span $\langle R\,\rangle$ meets the closed fundamental Weyl chamber
$\overline{K(S)}$ in exactly one point. Taken together, the following
map $\psi$ from the set of $W$-orbits of $\Pi$-systems into
$\overline{K(S)}\cap{\rm Span}_{\cal Z}(R^\ast)$ is thus well defined:
\be
\psi(W\cdot\Sigma):=\mbox{point of the orbit of $\kappa(\Sigma)$ in
$\overline{K(S)}$}.
\lb{nsb3}
\ee

Now we can formulate the announced correspondence between $\Pi$-systems
and $A_1$-vectors in $K(S)$:
\begin{theorem} With the notation introduced obove we have
\be
\mbox{\rm Im}(\psi)\cap K(S)=\big\{ \mbox{$A_1$-vectors in $K(S)$}
\big\}.
\lb{nsb4}
\ee
\end{theorem}

Note that this theorem implies also that the set on the right hand side
of (\ref{nsb4}) is determined entirely by the root system $R$.

\noindent{\sc Proof}{\hspace{0.75em}} We show first that
$\kappa(\Sigma)$ is an $A_1$-vector of $LT$ for any $\Pi$-system
$\Sigma$. {}From what has been done in Appendix A it is easy to guess
the
elements $\Lambda_\pm$ which lead to the correct commutation relations
(\ref{nsa12}). If we express $H:=\kappa(\Sigma)$ in the form
\[
\kappa(\Sigma)=\sum_{\alpha\in \Sigma}n(\Sigma)_\alpha
\frac{2\alpha}{\langle\alpha,\alpha\rangle}\,,
\]
then $\Lambda_\pm$ can be chosen as (compare with (\ref{nsa13}) and
(\ref{nsa15}))
\[
\Lambda_+=\sum_{\alpha\in S(H)} w^\alpha e_\alpha,\qquad
\Lambda_-=c(\Lambda_+)
\]
with the normalization
\[
{|w^\alpha |}^2 = \frac{1}{{(2\pi)}^2}\,
\frac{1}{\langle e_\alpha,c(e_\alpha)\rangle}\,
\frac{2n(\Sigma)_\alpha}{\langle\alpha,\alpha\rangle}\,.
\]
The orthogonality and commutation relations imply indeed that
(\ref{nsa12}) is fulfilled.

It remains to show that for each $A_1$-vector $H$ in $K(S)$ there
exists a $\Pi$-system $\Sigma$ with $\kappa(\Sigma)=H$. Now, we know
from the proof of Theorem A1 that $S(H)$, defined in (\ref{nsa1}), is a
$\Pi$-system, and eq.\ (\ref{nsa16}) tells us that
$\kappa(S(H))=2\varrho(H)^\ast=H$. This completes the proof of Theorem
B1.

One can actually show quite easily that $\Sigma=S(H)$ in the last step
is the only $\Pi$-system whose image is $H$.

At this point we could show that Theorem B1, together with Dynkin's
description of the $\Pi$-systems of semisimple Lie algebras, provides
an efficient method to determine all $A_1$-vectors in $K(S)$. The
details of this will be described elsewhere. As an example, the method
leads quickly to the following result for the Lie algebra $A_n$
corresponding to $SU(n+1)$: For $n$ {\em\/odd\/} there is only one
$A_1$-vector in the open fundamental Weyl chamber, namely the sum of
positive inverse roots $2\varrho^\ast=\sum_{\alpha\in R_+}
\alpha^\ast$. For $n$ {\em even} there are more possibilities,
corresponding to the following types of $\Pi$-systems: $A_n$,
$A_{n-1}$, $A_{n-2}+A_1$, $A_{n-3}+A_2,\ldots,A_{n/2}+A_{n/2-1}$.
\end{document}